\documentclass[twocolumn,showpacs,amsmath,amssymb,superscriptaddress]{revtex4-1}

\usepackage{graphicx}
\usepackage{dcolumn}
\usepackage{bm}
\usepackage[abs]{overpic}
\usepackage{here}
\usepackage{color}
\def\up{\uparrow}
\def\down{\downarrow }
\def\Vec#1{\bm{#1}}

\begin{document}


\title{
Critical Temperature Enhancement of Topological Superconductors: \\ A Dynamical Mean Field Study
}

\author{Yuki Nagai}
\affiliation{CCSE, Japan  Atomic Energy Agency, 178-4-4, Wakashiba, Kashiwa, Chiba, 277-0871, Japan}
\author{Shintaro Hoshino}
\affiliation{Department of Basic Science, The University of Tokyo, Meguro, Tokyo, 153-8902, Japan}
\author{Yukihiro Ota$^{1,}$}
\altaffiliation{
Present address: Research Organization for Information Science and Technology (RIST),
1-5-2 Minatojima-minamimachi, Kobe, 650-0047, Japan}


\date{\today}
             
\begin{abstract}
We show that a critical temperature $T_{\rm c}$ for spin-singlet two-dimensional superconductivity is enhanced by a 
cooperation between the Zeeman magnetic field and the Rashba spin-orbit coupling, where 
a superconductivity becomes topologically non-trivial below $T_{\rm c}$. 
The dynamical mean field theory (DMFT) with the segment-based hybridization-expansion continuous-time quantum Monte Carlo impurity solver (ct-HYB) is used for accurately evaluating a critical temperature, without any Fermion sign problem.
A strong-coupling approach shows that spin-flip driven local pair hopping leads to part of this enhancement, especially effects of the magnetic field.
We propose physical settings suitable for verifying the present calculations, one-atom-layer system on Si(111) and ionic-liquid based electric double-layer transistors (EDLTs). 
\end{abstract}

\pacs{
74.20.Rp, 
74.25.-q, 
74.25.Dw 
}
\maketitle

Interesting materials properties are produced by the interplay between
different internal degrees of freedom, such as spin and orbital, leading
to the design of devices with useful characteristics~\cite{Zelse,Hasan}.  
Manipulating spins in position or momentum space allows us to address
exotic order in low-temperature physics. 
The application of Zeeman magnetic fields induces a spin imbalance in a
system.  
Spin-orbit couplings (SOC) create a spin rotation depending on electron's
motion. 
These effects lead to notable many-body ground 
states, such as the Fulde-Ferrel-Larkin-Ovchinnikov
states~\cite{Fulde,Larkin,YanasePRB}, pair-density wave~\cite{Yoshida},
and topological
superfluidity/superconductivity~\cite{Kitaev,Nayak,Alicea}. 

The quest for high-$T_{\rm c}$ topological superconductors is a
compelling issue in materials science. 
To reveal a way of enhancing $T_{\rm c}$ with keeping topological
characters enables us to not only study topological order in a wide
range of temperatures but also increase the feasibility of implementing
topological quantum computing.
Superconducting topological insulator
$\mbox{Cu}_{x}\mbox{Bi}_{2}\mbox{Se}_{3}$ shows superconductivity at
$T_{\rm c} \sim 3.8\,\mbox{K}$~\cite{Sasaki}, and is a candidate for
bulk topological superconductors~\cite{Fu,Sasaki}. 
Interestingly, its critical temperature is two orders of magnitude
larger than a theoretical estimation with electron-phonon
couplings~\cite{ZhangSci}. 
Therefore, using a concrete theoretical method beyond the weak-coupling
mean-field theory, clarifying the relevance of the key features of this
compounds to $T_{\rm c}$ would lead to a clue of designing useful
topological materials. 

The presence of strong SOC is one of the crucial characters in  $\mbox{Cu}_{x}\mbox{Bi}_{2}\mbox{Se}_{3}$, since 
the quasiparticle wavefunction has a strong momentum dependence due to the SOC so that the bulk state has a nontrivial topology\cite{Fu}. 
This feature is common with other topological superconducting systems,
such as ultra-cold atomic gases and artificial
semiconductor-superconductor hetero structures~\cite{Sau}. 
Thus, it is interesting how spin degrees of freedom contribute to
the critical temperature of topological superconductors. 

\begin{figure}[t]
\begin{center}
     \begin{tabular}{p{ 0.8 \columnwidth}} 
      \resizebox{0.8 \columnwidth}{!}{\includegraphics{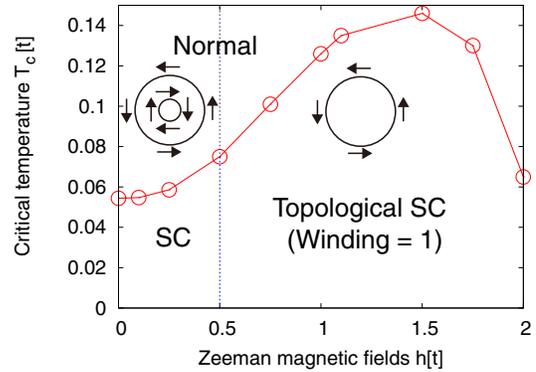}} 
    \end{tabular}
\end{center}
\caption{(Color online) Zeeman-magnetic-field dependence of a critical temperature, with the attractive on-site coupling $U = -3t$, the spin-orbit coupling $\alpha = 1t$, the filling $\nu = 1/8$. 
The dash-dotted line shows that in a non-interacting case ($U=0$) a winding number on the Fermi surfaces changes. 
\label{fig:fig1}
 }
\end{figure}

An attractive idea of producing topological superconductors is
to use 2D $s$-wave superconductors with spin manipulations~\cite{Sato}. 
Mean-field calculations predict that spin-singlet Cooper pairs have
unconventional and topological characters in the presence of Rashba SOC
and Zeeman magnetic fields~\cite{Sato,Shitade,NagaiJPSJ}. 
This setting is suitable for assessing the connection between
SOC and $T_{\rm c}$, from two points of view.
First, an intrinsic pair breaking (Pauli depairing) effect is involved,
owing to the presence of Zeeman magnetic fields.
Thus, one may understand how the contributions from SOC
overcome the Pauli depairing effects. 
Second, a topologically non-trivial $s$-wave superconducting state is
produced by an on-site attractive density-density
interaction~\cite{Sato}.
It indicates that an arbitrary range of interaction strength can be 
systematically studied by a reliable theoretical method, the dynamical
mean field theory (DMFT)~\cite{Georges:1996hv} combined with a
numerically exact continuous-time quantum Monte Carlo method
\cite{Werner,Gull}.
Utilizing this theoretical approach, one can take all kinds of local
Feynman diagrams.

\begin{figure}[t]
\begin{center}
     \begin{tabular}{p{ 0.8 \columnwidth}} 
      (a)\resizebox{0.8 \columnwidth}{!}{\includegraphics{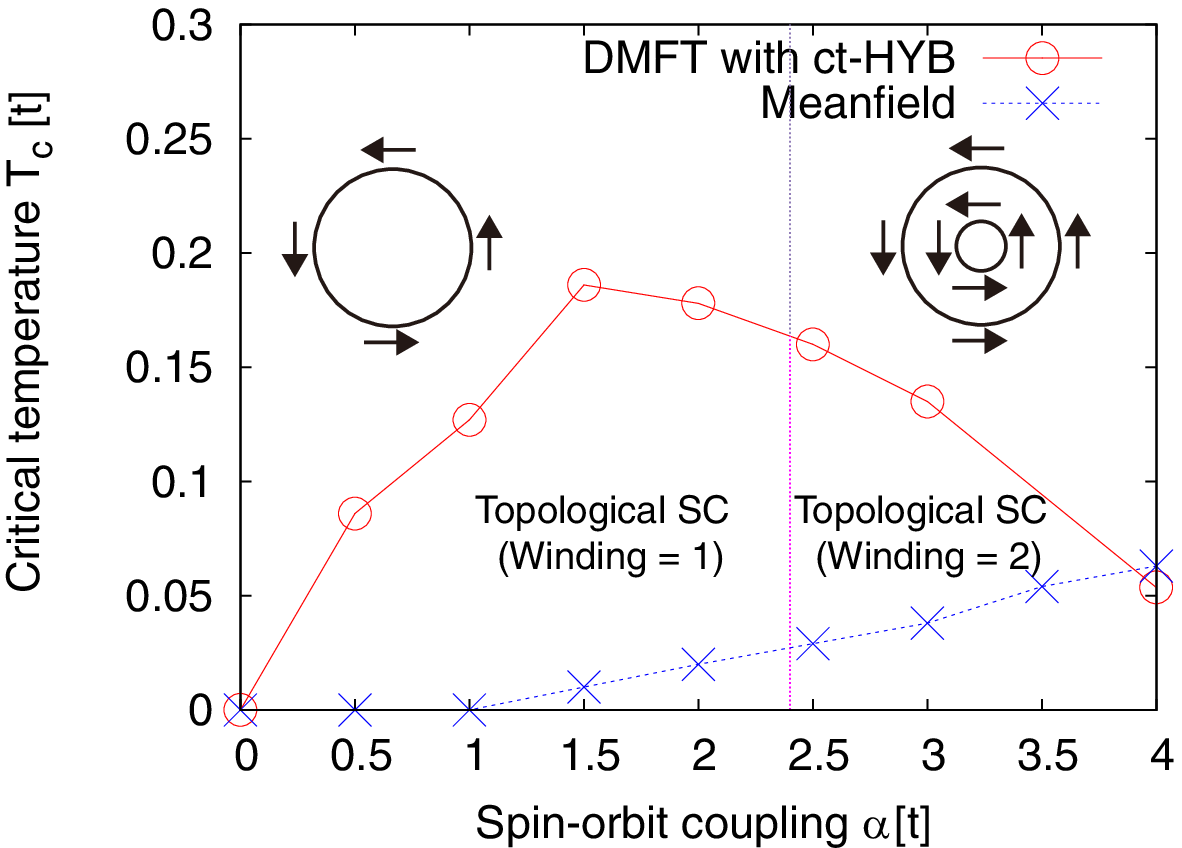}} \\
      (b)\resizebox{0.8 \columnwidth}{!}{\includegraphics{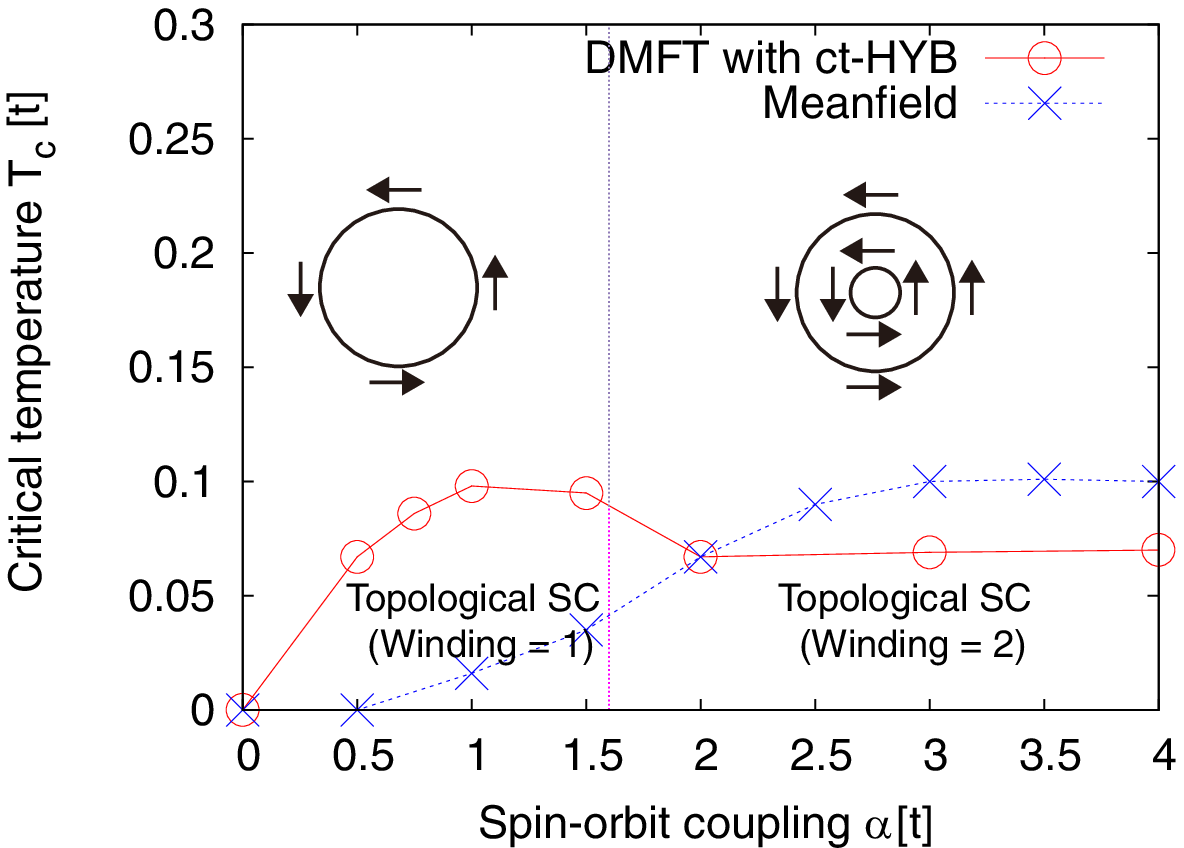}} 
    \end{tabular}
\end{center}
\caption{
\label{fig:fig2}(Color online) Spin-orbit coupling dependence of a critical 
temperature, with (a) Zeeman magnetic field, $h=t$ and filling, $\nu=1/8$ and (b) $h=0.5t$ and $\nu=1/16$. The attractive on-site coupling, $U=-3t$, is common. 
Each vertical dash-dotted line has the same meaning as that in Fig.~\ref{fig:fig1}. 
%
 }
\end{figure}

In this paper, we show that a 2D attractive Hubbard
model with Rashba SOC possesses $T_{\rm c}$ enhancement even though the
Zeeman magnetic field is applied.
To treat an arbitrary strength of on-site $U$, we adopt the DMFT
combined with a numerically exact continuous-time quantum Monte Carlo
method. 
We point out that this approach accurately estimates $T_{\rm c}$ even in
the present spin-active many-body system since a symmetric
property of the many-body Hamiltonian in spin and $k$-space ensures the absent of the Fermion
negative sign problem\cite{Gull}.  
Our main results are shown in Fig.~\ref{fig:fig1} and Fig.~\ref{fig:fig2}.
The critical temperature on a certain filling 
changes with a non-monotonic manner, varying the magnitude of the Zeeman field and the Rashba
SOC.
A cooperation between the Zeeman field and SOC is a key of the $T_{\rm c}$ enhancement. 
A strong-coupling approach shows that the part of the enhancement (i.e. the magnetic field dependence) is explained by the local pair hopping~\cite{Micnas:1990ee}, due to a spin-flip process
%
The rest of the enhancement (i.e., the SOC dependence) is still elusive. 
We speculate that the enhancement is related to a change of a winding number on the normal-electron Fermi surfaces.
Moreover, we propose physical settings suitable for verifying the
present calculations, one-atom-layer \mbox{TI-Pb} on
\mbox{Si(111)}\cite{Matetskiy} and ionic-liquid based electric
double-layer transistors (EDLTs)\cite{Li}.

The single-orbital attractive Hubbard Hamiltonian 
with the Rashba SOC and the Zeeman magnetic field on 2D square lattice is\cite{Sau,Sato}  
\begin{align}
{\cal H} &= \sum_{\Vec{k} \sigma \sigma'} \hat{h}_0^{\sigma \sigma'}(\Vec{k}) c_{\Vec{k} \sigma}^{\dagger} c_{\Vec{k} \sigma'} + 
U \sum_{i} n_{i \up} n_{i \down} ,
\end{align}
where 
\(
\hat{h}_{0}(\Vec{k}) 
= 
-\mu - 2 t(\cos k_{x} + \cos k_{y})
+ \alpha {\cal L}(\Vec{k})
- h \hat{\sigma}_{3}
\) 
and 
$n_{i \sigma} = c_{i \sigma}^{\dagger}c_{i \sigma}$ 
($\sigma=\up,\,\down$). 
The hopping parameter, $t$ is positive, whereas the coupling constant
of the on-site interaction, $U$ is negative.
Throughout this paper, we use the unit system with $\hbar=k_{\rm B}=1$.
The unit of energy is $t$. 
The electron annihilation (creation) operator with spin
$\sigma$ is $c_{i \sigma}$ ($c_{i \sigma}^{\dagger}$) on spatial site $i$. 
In the momentum representation they are 
$c_{\Vec{k} \sigma}$ and $c_{\Vec{k} \sigma}^{\dagger}$. 
The symbol $\hat{\sigma}_{j}$ is the $j$th component of the $2 \times 2$
Pauli matrices ($j = 1,2,3$). 
The Rashba SOC term is described by 
$\alpha {\cal L}(\Vec{k}) = \alpha (\hat{\sigma}_{1} \sin k_{y} - \hat{\sigma}_{2} \sin k_{x})$, 
with positive $\alpha$. 
The strength of the Zeeman magnetic field is $h$. 
In our calculations, the chemical potential, $\mu$ is tuned, with fixed
filling, $\nu$.

Let us summarize the topological properties of a superconducting
state in this model within the weak-coupling Bardeen-Cooper-Schrieffer
(BCS) theory~\cite{Sato}. 
The topological number is the Thouless-Kohmoto-Nightingale-Nijs
invariant\,\cite{Thouless,Kohmoto} on a 2D torus in the momentum space. 
According to this invariant, the criteria of topological
superconductivity are derived by Sato \textit{et al.} (Table I in
Ref.~\cite{Sato}).
We focus on the case just below $T_{\rm c}$; the amplitude of the
superconducting order parameter vanishes
(i.e. $|\Delta| \rightarrow 0+$).
Then, we find that the criteria in Ref.~\cite{Sato} are regarded as the
changes of a winding number on the Fermi surfaces, where the winding
number is defined as $xy$-plane spin rotation on the Fermi surfaces.
Note that this characterization requires
only the knowledge on the normal-state Fermi surfaces.
The occurrence of a topological superconducting state just
below $T_{\rm c}$ is associated with a non-zero winding number on the
Fermi surfaces. 
Hence, although in this paper we only consider the normal states
just above $T_{\rm c}$, we can obtain a connection of normal-state
instability with topological superconductivity.

We show our calculation method. 
To calculate one- and two-particle Green's functions, we utilize the
DMFT with the segment-based hybridization-expansion continuous-time
quantum Monte Carlo impurity solver (ct-HYB)~\cite{Werner,Haule,WernerPRL}.
The segment-based algorithm is the fastest update method of ct-HYB
solvers, 
and is applicable to our system if 
(i) the interaction terms of the
Hamiltonian conserve spin and (ii) in the effective Anderson impurity
model the one-body local Hamiltonian does so. 
The first condition is satisfied since the system has only density-density
interaction.
Let us consider the second one.
The one-body local Hamiltonian matrix, $\hat{H}_{\rm f}$, is related to
$\hat{h}_{0}(\Vec{k})$ via\,\cite{note} 
\begin{align}
\hat{H}_{\rm f} &= \sum_{\Vec{k}} \hat{h}_{0}(\Vec{k}) = - \mu  -
 h \hat{\sigma}_{3}. \label{eq:hf}
\end{align}
Since $\hat{H}_{\rm f}$ is diagonal in the spin space, the second
condition is fulfilled. 
Moreover, we point out that there is no Fermion sign problem
when self energy is diagonal in the spin space\cite{Gull}.  
Since ${\cal H}$ is invariant
under the transformation
$c_{\Vec{k} \sigma} \rightarrow \sum_{\sigma'}
(\hat{\sigma}_{3})_{\sigma \sigma'} c_{-\Vec{k} \sigma'}$, 
we find that the off-diagonal elements of self energy in the spin space
are zero in the present system~\cite{note2}.
Accordingly, the evaluation of $T_{\rm c}$ is accurately performed by
the DMFT with ct-HYB. 
In this paper, the effective impurity problem is solved by 
an open-source program package, {\it i}Qist~\cite{iQist}.

The main target in our calculations is the pair susceptibility with
respect to a spin-singlet $s$-wave state at temperature $T$~\cite{Hoshino},  
\begin{align}
\chi &= \frac{1}{N} \int_{0}^{1/T} \langle 
{\cal O}(\tau) {\cal O}^{\dagger}
\rangle d\tau = T \sum_{n n'} \chi_{\up \down \down \up} (i \omega_{n},i\omega_{n'};0), \label{eq:sus}
\end{align}
with ${\cal O} = \sum_{i} c^{\dagger}_{i \up} c^{\dagger}_{i \down}$.  
The total number of lattice sites is $N$.  
The fermionic Matsubara frequency is $\omega_{n} = \pi T (2n +1)$, with
$n \in \mathbb{Z}$. 
Here, $\chi_{abcd}(i \omega_{n},i\omega_{n'};0)$ is a two-particle lattice Green's function with a zero Bosonic Matsubara frequency.
A divergence in $\chi$ (or equivalently a sign change in $1/\chi$) indicates a possible transition into a superconducting phase. 
In the effective impurity model one- and two- particle
local Green's functions,
$G_{ab}^{\rm loc}(i \omega_{n})$ and
$\chi_{abcd}^{\rm loc}(i \omega_{n},i \omega_{n'};0)$ respectively,
are calculated by the ${\cal G}$ardenia component of the {\it i}Qist
package.
One-particle Green's function in the original lattice model is
$\hat{G}(\Vec{k}, i \omega_{n}) \equiv [i \omega_{n} - \hat{h}_{0}(\Vec{k}) - \hat{\Sigma}(i \omega_{n})]^{-1}$, 
with self energy $\hat{\Sigma}(i \omega_{n})$. 
Two-particle lattice Green's functions are obtained by
simultaneously solving two Bethe-Salpeter equations with a common vertex
function 
$\underline{\underline{\Gamma}}$~\cite{Hoshino,Freericks},
\begin{subequations}
\begin{align}
& \underline{\underline{\chi}}^{\rm loc} =
 \underline{\underline{\tilde{\chi}}}^{{\rm loc},0} +
 \underline{\underline{\chi}}^{{\rm loc},0} \:
 \underline{\underline{\Gamma}} \:\underline{\underline{\chi}}^{\rm loc}
 , \\
& \underline{\underline{\chi}} =
 \underline{\underline{\tilde{\chi}}}^{0}+
 \underline{\underline{\chi}}^{0} \: \underline{\underline{\Gamma}} \:
 \underline{\underline{\chi}} .
\end{align}
\end{subequations}
The double underline indicates that an object is a matrix
on a vector space including two spin indices and the Matsubara frequency;
$\chi_{abcd}(i \omega_{n},i \omega_{n'})$ is embedded into
\(
(\underline{\underline{\chi}} )_{l l^{\prime}}
\)
with $l=(a,b,n)$ and $l'=(d,c,n')$, for example. 
We take all the processes of the DMFT framework, regardless of spin conservation or not.
In the Bethe-Salpeter equations the matrix objects with superscript $0$
contain bare two-particle Green's functions produced by one-particle
Green's functions.
In the effective impurity model, we have
\(
\chi_{abcd}^{{\rm loc},0}(i\omega_{n},i\omega_{n'})
= \chi_{dacb}^{{\rm loc},gg}(i \omega_{n}) \delta_{n,n'}
\)
and
\(
\tilde{\chi}_{abcd}^{{\rm loc},0}(i\omega_{n}, i\omega_{n'})
=
\chi_{abcd}^{{\rm loc},0}(i\omega_{n}, i\omega_{n'})
-
\chi_{cadb}^{{\rm loc},gg}(i \omega_{n}) \delta_{n,-n'-1}
\),
with
\(
\chi_{abcd}^{{\rm loc},gg}(i \omega_{n})
= G_{ab}^{\rm loc}(i \omega_{n})
G_{cd}^{\rm loc}(- i \omega_{n}) 
\). 
In a similar manner we define bare two-particle Green's functions in the
lattice model; 
all local one-particle Green's functions are replaced with lattice
one-particle Green's functions, and $\chi_{abcd}^{gg}(i \omega_{n})$ is
defined as 
\(
\chi_{abcd}^{gg}(i \omega_{n})
= \sum_{\Vec{k}}
G_{ab}(\Vec{k}, i \omega_{n})G_{cd}(-\Vec{k},- i \omega_{n}) 
\).
In the calculation of the two-particle Green's functions, the
$\Vec{k}$-mesh size and the $n$-mesh size are $192 \times 192$ and $64$,
respectively.
The numerical calculations on $\underline{\underline{\chi}}$ are perfomed by an equation not explicitly including $\underline{\underline{\Gamma}}$: 
$\underline{\underline{\chi}} = \underline{\underline{\chi}}^{\rm loc} (\underline{\underline{1}} - \underline{\underline{A}})^{-1} \underline{\underline{B}}$ 
with $\underline{\underline{A}} \equiv ([\underline{\underline{\chi}}^{\rm loc,0}]^{-1} -[\underline{\underline{\chi}}^{0}]^{-1}) \underline{\underline{\chi}}^{\rm loc} + \underline{\underline{1}} - 
\underline{\underline{B}}^{\rm loc}$, $\underline{\underline{B}} \equiv [\underline{\underline{\chi}}^{0}]^{-1} \underline{\underline{\tilde{\chi}}}^{0}$ and $\underline{\underline{B}}^{\rm loc} \equiv 
[\underline{\underline{\chi}}^{{\rm loc},0}]^{-1} \underline{\underline{\tilde{\chi}}}^{\rm loc,0}$ (in detail, see Ref.~\cite{suppl}).

Figure \ref{fig:fig1} shows $T_{\rm c}$ with respect to the change of the Zeeman magnetic field when $\alpha = t$, $U = -3t$, and
$\nu = 1/8$. 
The critical temperature increases with increasing $h$, and takes a peak around $h = 1.5 t$. 
Then, the decrease of $T_{\rm c}$ occurs in a stronger magnetic field; this reduction corresponds to the Pauli depairing effect. 
We mention that the weak-coupling mean-field calculations indicate the complete suppression of $T_{\rm c}$ even in the weak magnetic field $h = 1 t$ [See, e.g., Fig.~\ref{fig:fig2}(a) at $\alpha = 1 t$]. 
We note that a weak-coupling approach in the presence of spatial phase fluctuations \cite{Xu,Devreese} predicts the decrease of Tc with increasing Zeeman magnetic fields. 
%
We infer from Fig. 1 a relation between the $T_{\rm c}$ enhancement and 
the change of the winding number on the
Fermi surfaces from $0$ (conventional, non-topological, $s$-wave) to $1$ (topological $s$-wave). 
To study this point more closely, we consider a different way of changing the winding number. 
We focus on a region of parameter sets in which the winding number transits from $1$ to $2$ increasing $\alpha$ with fixed $\nu$ and $h$.  
Figure \ref{fig:fig2} shows the behaviors of $T_{\rm c}$ in the DMFT calculations (red circle), as well as the results obtained by the weak-coupling mean-field calculations (blue cross). 
We find in the DMFT calculations that the behavior of $T_{\rm c}$ is non-monotonic as $\alpha$. 
In contrast, the weak-coupling mean-field critical temperature monotonically grows up as $\alpha$ since the in-plane Rashba SOC may suppress the Pauli depairing effect induced by the Zeeman magnetic field along $z$-axis~\cite{note3}. 
In the DMFT calculations, an optimal value of $\alpha$ in the enhancement of $T_{\rm c}$ locates at the region of the winding number to be $1$.
Thus, our calculations suggest that the parameter region in which the winding number is $1$ be suitable for realizing a topological superconducting state at a high temperature.
We stress that these results occur at different parameter sets, as shown in Figs.~\ref{fig:fig2}(a) and (b).
It is important to note that 
calculating a topological invariant in interacting systems is desirable for finding the genuine topological transition point. 
The renormalized Zeeman magnetic field and the chemical potential due to the self-energy at the zero-energy \cite{Wang:2012fjc} would shift the lines of winding-number changes in 
Figs.~\ref{fig:fig1} and \ref{fig:fig2}.

Now, we derive a strong-coupling-limit formula of $T_{\rm c}$, to get the picture on the enhancement of $T_{\rm c}$ with respect to the changes of $\alpha$ and $h$. 
In a strong-coupling limit $|U| \rightarrow \infty$, the model can be rewritten as a pseudospin $(S=1/2)$ quantum Heisenberg model, whose Hamiltonian is given by 
${\cal H}_{\rm eff} = \sum_{\langle i j \rangle} [-J (S_{i}^{x} S_{j}^{x} + S_{i}^{y} S_{j}^{y}) + J S_{i}^{z} S_{j}^{z}] -H \sum_{i} S_{i}^{z}$, where $\sum_{\langle i j \rangle}$ considers 
nearest-neighbors only and the pseudospin up and down states are doubly-occupied and unoccupied local states, respectively. 
The effective coupling constant and the pseudo spin field are $J(t,\alpha,h,U) \equiv 4 t^{2}/|U| +  |U| \alpha^{2}/(|U|^{2}-4 h^{2})$ and $H = - 2 \mu + U$, respectively. 
The mean-field analysis in this effective Hamiltonian gives the critical temperature. 
Thus, we obtain the expression of
$T_{\rm c}$, 
\begin{align}
T_{\rm c}(\alpha,h)  &= \cfrac{ (1 -2 \nu) J(t,\alpha,h,U) }{ \tanh^{-1} \left(1 - 2 \nu \right)}.
  \label{eq:strong}
\end{align}

\begin{figure}[t]
\begin{center}
     \begin{tabular}{p{ 0.8 \columnwidth}} 
      \resizebox{0.8 \columnwidth}{!}{\includegraphics{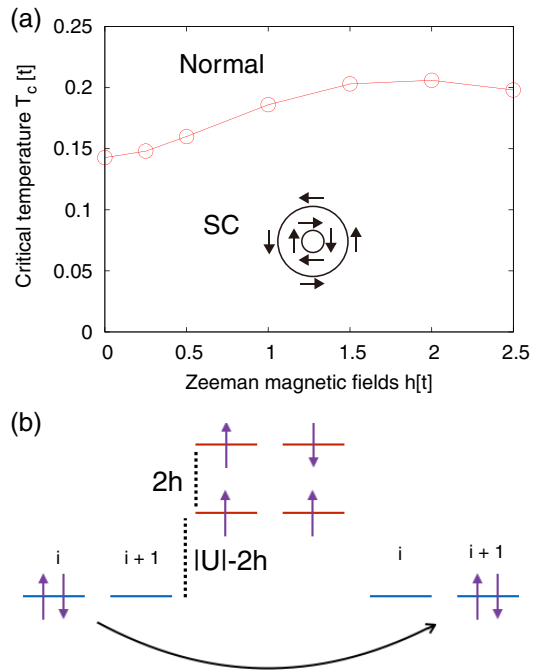}} 
    \end{tabular}
\end{center}
\caption{(Color online) (a) Zeeman magnetic field dependence of a critical temperature at half filling ($\nu=1/2$). Other settings are equal to those in Fig.~\ref{fig:fig1}.
%
(b) 
Schematic diagram of pair hopping from the $i$th to $(i+1)$th sites via either spin-flip (lower panel) or spin-conserved (upper panel) processes. 
\label{fig:fig3}
 }
\end{figure}

We show that part of the enhancement ($h$-dependence of $T_{\rm c}$) is explained by local pair hopping~\cite{Micnas:1990ee} in terms of the strong coupling approach. 
We find that, even in the half filling case ($\nu = 1/2$), 
the critical temperature enhances with increasing $h$, as shown in Fig.~\ref{fig:fig3}(a). 
A key of the enhancement of $T_{\rm c}$ is depicted in Fig.~\ref{fig:fig3}(b); a pair on the $i$th site can hop into the
$i + 1$th site via a virtual spin-flip process coming from non-vanishing $\alpha$ of second-order perturbation (lower diagram on the middle panel).
A strong Zeeman magnetic field splits the energy levels between the spin-flip (lower diagram) and spin-conserved (upper diagram) processes; 
the presence of the Zeeman field tends to increase the rate of the spin-flip process. 
Therefore, 
this spin-flip-driven local pair hopping is responsible for the $T_{\rm c}$ enhancement under nonzero $h$, although most of our DMFT calculations are outside strong-coupling regime since the energy of the singly-occupied  state is smaller than the energies of the doubly-occupied and empty states.
In Ref.~\cite{suppl}, we also show  that the spin-flip processes are important for the $T_{\rm c}$ enhancement in the DMFT calculations.
%

The local-pair-hopping scenario, however, does not fully explain the behavior of $T_{\rm c}$. 
Under the fixed Zeeman magnetic field, the dependence of $T_{\rm c}$ on $\alpha$ in the strong-coupling formula is quite
different from the DMFT calculations, as shown in Fig.~\ref{fig:fig2}; the critical temperature in the DMFT calculations is {\it not} a monotonic increase function of
$\alpha$, even though the Pauli depairing effect could be suppressed for large $\alpha$. 
We can find that the DMFT calculations with $U=-7\,t$ are consistent with the strong-coupling formula; $T_{\rm c}$ monotonically increases within our calculations in $0 \le \alpha \le 4\,t$.
Thus, explaining the $\alpha$-dependence of $T_{\rm c}$ in an intermediate range of on-site $U$ would require for a different scenario. 
It is an interesting future issue of unveiling the remaining origin of $T_{\rm c}$ enhancement. 
We speculate that focusing on the change of the winding number might give us an insight on this elusive issue.

\begin{figure}[t]
\begin{center}
     \begin{tabular}{p{ 0.8 \columnwidth}} 
      \resizebox{0.8 \columnwidth}{!}{\includegraphics{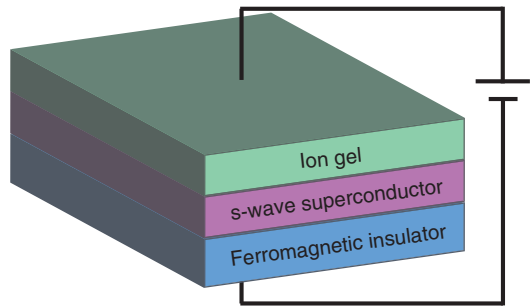}} 
    \end{tabular}
\end{center}
\caption{
\label{fig:fig4}(Color online)  Schematic figure of ionic-liquid based electric double-layer transistors (EDLTs).
 }
\end{figure}

Now, we propose two physical systems available for testing our theoretical calculations. 
The first setup is to apply Zeeman magnetic fields to one-atom-layer Tl-Pb compounds on Si(111).
Matetskiy \textit{et al.}~\cite{Matetskiy} observed the occurrence of giant Rashba effects in this setting without the Zeeman field. 
The second setup is to use EDLT with a layered structure built up by an $s$-wave superconductor and a ferromagnetic insulator, as shown in
Fig.~\ref{fig:fig4}. 
The idea of fabricating related systems is shown in Fig.~2(a) of Ref.~\cite{Li}.  
Tuning electric fields allows us to control electron's filling and the strength of SOC. 
This EDLT setup would be a plausible system to design 2D topological superconductors.

Finally, we discuss a link of our calculations with Cu$_{x}$Bi$_{2}$Se$_{3}$, from the viewpoint of $T_{\rm c}$ enhancement. 
Although the present system is quite different from $\mbox{Cu}_{x}\mbox{Bi}_{2}\mbox{Se}_{3}$, 
we have an interesting correspondence between the two systems. 
One of the authors (Y.N.) found in the calculations of impurity effects~\cite{NagaiRapid} that the presence of orbital imbalance leads
to similar effects to those induced by a spin imbalance, even though an
external magnetic field is absent. 
In the present system a strong Zeeman magnetic field
induces the coherent hopping of a localized pair via a
spin-flip process, leading of the increase of $T_{\rm c}$. 
Hence, in $\mbox{Cu}_{x}\mbox{Bi}_{2}\mbox{Se}_{3}$, a large orbital
imbalance might cause an orbital-flip process contributing to the
enhancement of $T_{\rm c}$. 
The DMFT study in the model of Cu$_{x}$Bi$_{2}$Se$_{3}$ is our
important future issue.

In summary, 
we showed that a 2D attractive Hubbard model with Rashba SOC and a Zeeman magnetic field  possesses $T_{\rm c}$ enhancement,  
by using the DMFT combined with the numerically exact ct-HYB solver without any Fermion sign problem. 
With the use of a strong-coupling approximation, part of the enhancement (i.e.~the magnetic field dependence) was explained by 
the scenario of a local pair hopping induced by a spin-flip process. 
The rest of the enhancement (i.e.~the SOC dependence) is still in an open issue. 
We speculated that the enhancement is related to a change of a winding number of the normal-electron Fermi surfaces.
Moreover, we proposed that EDLTs are good stages for designing topological superconductivity. 
Finally, we discussed a high $T_{\rm c}$ in Cu$_{x}$Bi$_{2}$Se$_{3}$ with the use of the result in our two-dimensional system.

Y. N. thanks Y. Saito for helpful comments on the EDLTs. 
The calculations were performed by the supercomputing 
system SGI ICE X at the Japan Atomic Energy Agency. 
This study was partially supported by JSPS KAKENHI Grants No. 26800197 and No. 15K00178.


\clearpage

\onecolumngrid
\setcounter{equation}{0}
\renewcommand{\thefigure}{S\arabic{figure}} 

\setcounter{figure}{0}

\renewcommand{\thesection}{S\arabic{section}.} 
\renewcommand{\theequation}{S\arabic{equation}} 
\begin{flushleft} 
{\Large {\bf Supplemental materials}}
\end{flushleft} 
\begin{flushleft} 
{\bf S1. Two Bethe-Salpeter equations}
\end{flushleft} 
This section closely shows a heart of our formulation in a system with spin-orbital couplings. 
The pair susceptibility with respect to a spin-singlet $s$-wave state $\chi$ is  Eq.~(3) in the main text. A divergence in $\chi$ (or equivalently a sign change in $1/\chi$) indicates a possible transition into a superconducting phase. 
This quantity is obtained by simultaneously solving two Bethe-Salpeter equations, one of which is formulated in an effective impurity model, while another of which does in the original lattice model.
The Bethe-Salpeter equation for the effective impurity model is expressed as 
\begin{align}
\chi_{aa'a'a}^{{\rm loc}}(i \omega_{n},i \omega_{n'}) 
&=
\left( 
 \chi_{a a a' a'}^{{\rm loc},0 }(i \omega_{n}) \delta(\omega_{n} - \omega_{n'}) 
 -\chi_{a' a a a'}^{{\rm loc},0 }(i \omega_{n}) \delta(\omega_{n} + \omega_{n'}) \delta_{a' a}
\right) \nonumber \\
&
+
\sum_{n_{1},n_{2}}
\left[
 \chi_{a a a' a'}^{{\rm loc},0 }(i \omega_{n}) \delta(\omega_{n} - \omega_{n_{1}})
 \right] 
\Gamma_{a'aaa'}(i \omega_{n_{1}},i\omega_{n_{2}}) \chi_{aa'a'a}^{{\rm loc}}(i \omega_{n_{2}},i \omega_{n'}). 
\label{eq:impb}
\end{align}
The subscripts ($a$, $a^{\prime}$, and so on) represent spin indices, equal to those in the main text.  
Here, we assume the spin-diagonal one-body local Hamiltonian [See Eq.~(2) in the main text] and the density-density interaction. 
The bare local two-particle Green's function $\chi_{abcd}^{{\rm loc},0 }(i \omega_{n})$ is defined by 
\begin{align}
\chi_{abcd}^{{\rm loc},0 }(i \omega_{n}) &= G_{ab}^{\rm loc}(i \omega_{n}) G_{cd}^{\rm loc}(- i \omega_{n}),
\end{align}
with $\hat{G}^{\rm loc}(i \omega_{n}) \equiv \sum_{\Vec{k}} \hat{G}(\Vec{k},i \omega_{n}) = \sum_{\Vec{k}}[i \omega_{n} - \hat{h}_{0}(\Vec{k}) - \hat{\Sigma}(i \omega_{n})]^{-1}$ calculated by an impurity solver.
Note that Eq.~(\ref{eq:impb}) can be solved with fixed indices $(a,a')$, separately. 
Thus, one can obtain $\chi_{\up \up \up \up}^{{\rm loc}}(i \omega_{n},i \omega_{n'})$, $\chi_{\up \down \down \up}^{{\rm loc}}(i \omega_{n},i \omega_{n'})$, 
$\chi_{\down \up \up \down}^{{\rm loc}}(i \omega_{n},i \omega_{n'})$, and $\chi_{\down \down \down \down}^{{\rm loc}}(i \omega_{n},i \omega_{n'})$, seperately,  
with the use of the impurity solver for the system where the spin is conserved. 
%
On the other hand, the Bethe-Salpeter equation for the original lattice model is expressed as 
\begin{align}
\chi_{aa'a'a}(i \omega_{n},i \omega_{n'}) 
&=
\left( 
 \chi_{a a a' a'}^{0 }(i \omega_{n}) \delta(\omega_{n} - \omega_{n'}) 
 -\chi_{a' a a a'}^{0 }(i \omega_{n}) \delta(\omega_{n} + \omega_{n'}) 
\right) \nonumber \\
&
+
\sum_{n_{1},n_{2}} \sum_{a_{1},a_{2},a_{3},a_{4}}
\left[
 \chi_{a_{2} a a_{1} a'}^{0 }(i \omega_{n}) \delta(\omega_{n} - \omega_{n_{1}})
 \right] 
\Gamma_{a_{1}a_{2}a_{3}a_{4}}(i \omega_{n_{1}},i\omega_{n_{2}}) \chi_{a_{3}a_{4}a'a}(i \omega_{n_{2}},i \omega_{n'}). 
\label{eq:latb}
\end{align}
Here, the vertex $\Gamma_{a_{1}a_{2}a_{3}a_{4}}(i \omega_{n_{1}},i\omega_{n_{2}})$ is equal to that in the local Bethe-Salpeter equation (\ref{eq:impb}). 
The bare original lattice two-particle Green's function $\chi_{abcd}^{0}(i \omega_{n})$ is defined by 
\begin{align}
\chi_{abcd}^{0}(i \omega_{n}) &= \sum_{\Vec{k}} G_{ab}(\Vec{k},i \omega_{n}) G_{cd}(-\Vec{k},- i \omega_{n}).
\end{align}

To solve two Bethe-Salpeter equations simultaneously, we introduce an expression $\underline{\underline{\chi}}$ on a vector space including two spin indices and the Matsubara frequency. 
$\chi_{abcd}(i \omega_{n},i \omega_{n'})$ is embedded into
\(
(\underline{\underline{\chi}} )_{l l^{\prime}}
\)
with $l=(a,b,n)$ and $l'=(d,c,n')$, for example. 
The numerical calculations on $\underline{\underline{\chi}}$ are performed by an equation not explicitly including $\underline{\underline{\Gamma}}$: 
\begin{align}
\underline{\underline{\chi}} &= \underline{\underline{\chi}}^{\rm loc} (\underline{\underline{1}} - \underline{\underline{A}})^{-1} \underline{\underline{B}}, 
\end{align}
with $\underline{\underline{A}} \equiv ([\underline{\underline{\chi}}^{\rm loc,0}]^{-1} -[\underline{\underline{\chi}}^{0}]^{-1}) \underline{\underline{\chi}}^{\rm loc} + \underline{\underline{1}} - 
\underline{\underline{B}}^{\rm loc}$, $\underline{\underline{B}} \equiv [\underline{\underline{\chi}}^{0}]^{-1} \underline{\underline{\tilde{\chi}}}^{0}$ and $\underline{\underline{B}}^{\rm loc} \equiv 
[\underline{\underline{\chi}}^{{\rm loc},0}]^{-1} \underline{\underline{\tilde{\chi}}}^{\rm loc,0}$.
The above expression is numerically stable, since the inverse matrices $[\underline{\underline{\chi}}^{\rm loc,0}]^{-1}$ and $[\underline{\underline{\chi}}^{0}]^{-1}$ are diagonal in the Matsubara space.

\begin{flushleft} 
{\bf S2. Comparison with full and spin-conserved processes}
\end{flushleft} 
In this section, we show that the spin-flip processes are important for the $T_{\rm c}$-enhancement in the main text. 
The full spin-singlet pairing susceptibility $\chi_{\up \down \down \up}(i \omega_{n},i \omega_{n'}) $ includes 
both spin-conserved and spin-flip processes. 
The Rashba spin-orbit coupling in Eq.~(1) in the main text induces spin-flipping processes. 
If the two-particle Green's function is constructed by the spin-conserved processes only, the Bethe-Salpeter equation (\ref{eq:latb}) in the original lattice system becomes 
\begin{align}
\chi_{\up \down \down \up}(i \omega_{n},i \omega_{n'}) 
&=
 \chi_{\up \up \down \down}^{0 }(i \omega_{n}) \delta(\omega_{n} - \omega_{n'}) 
+
\sum_{n_{1},n_{2}} 
\left[
 \chi_{\up \up \down \down}^{0 }(i \omega_{n}) \delta(\omega_{n} - \omega_{n_{1}})
 \right] 
\Gamma_{\down \up \up \down}(i \omega_{n_{1}},i\omega_{n_{2}}) \chi_{\up \down \down \up}(i \omega_{n_{2}},i \omega_{n'}). 
\label{eq:latbc}
\end{align}

\begin{figure}[bt]
\begin{center}
     \begin{tabular}{p{0.4 \columnwidth} p{0.4 \columnwidth} }
      (a)\resizebox{0.4 \columnwidth}{!}{\includegraphics{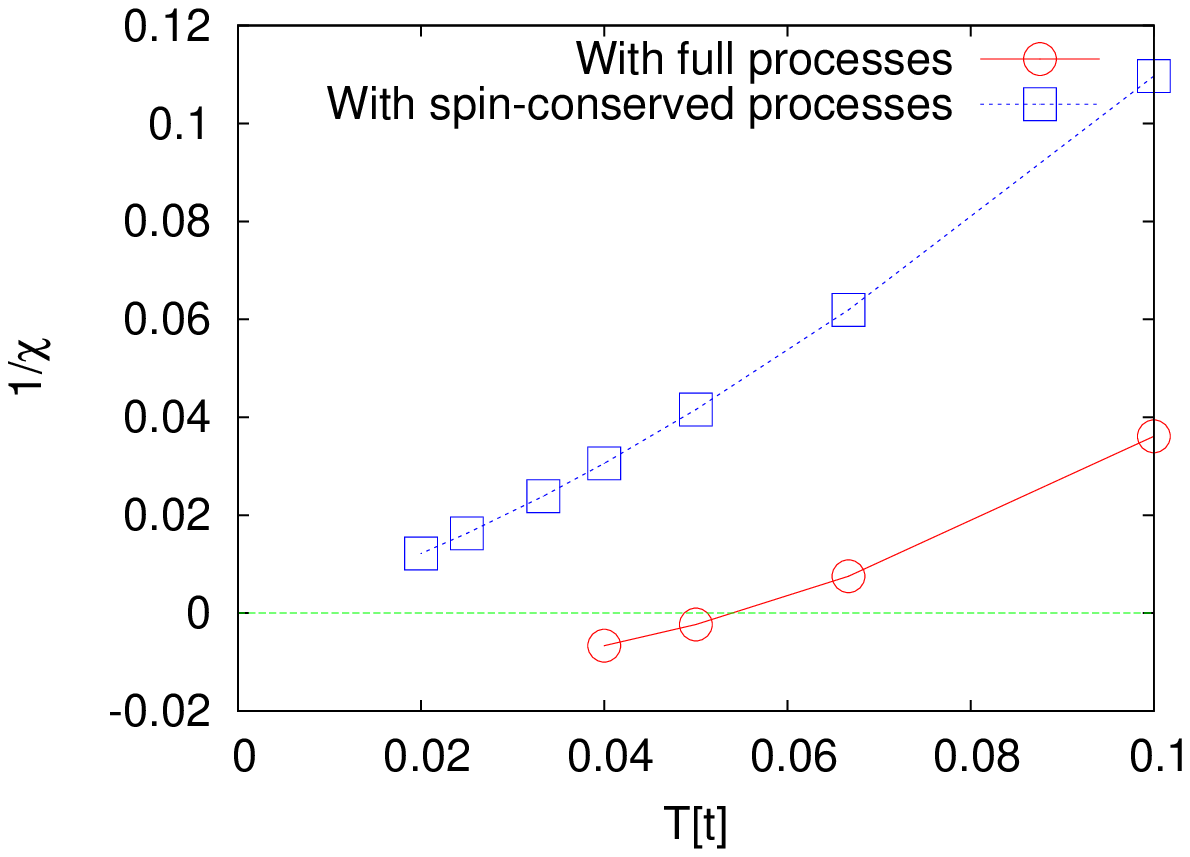}} &
      (b)\resizebox{0.4 \columnwidth}{!}{\includegraphics{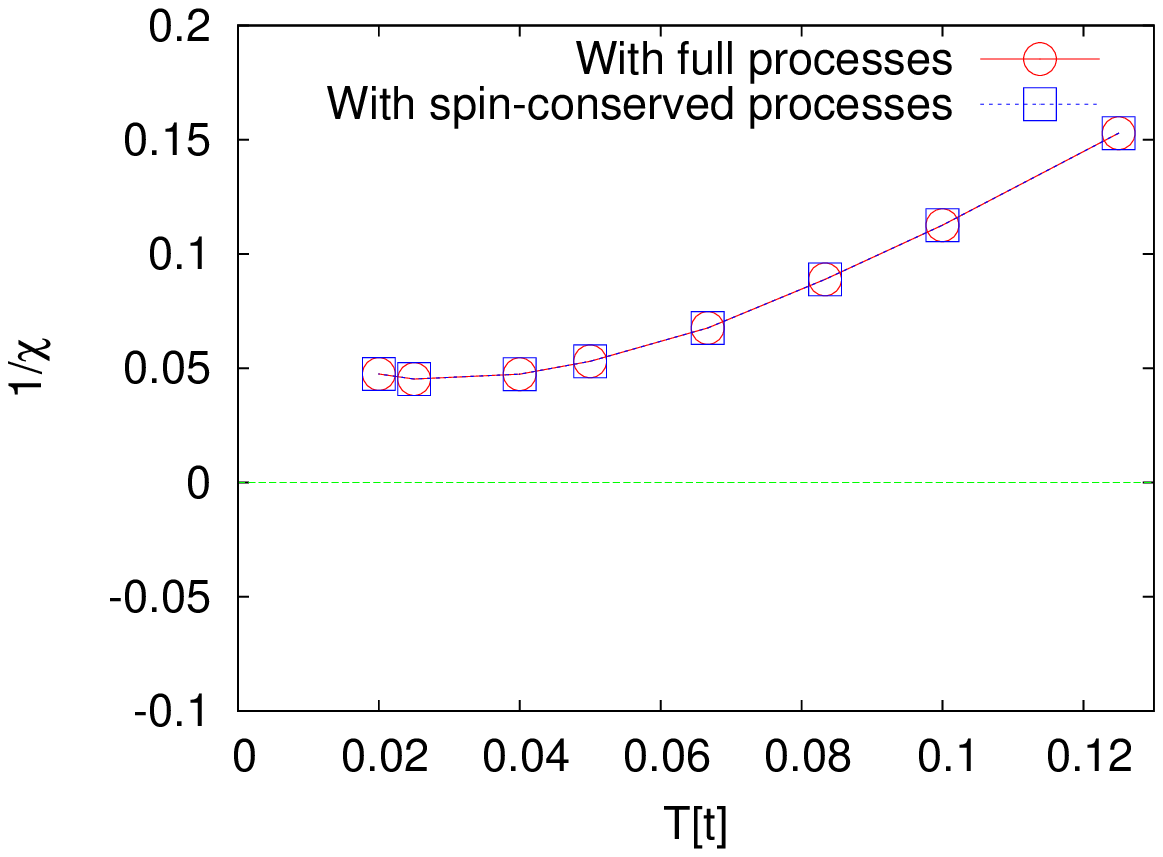}}  \\
       (c)\resizebox{0.4 \columnwidth}{!}{\includegraphics{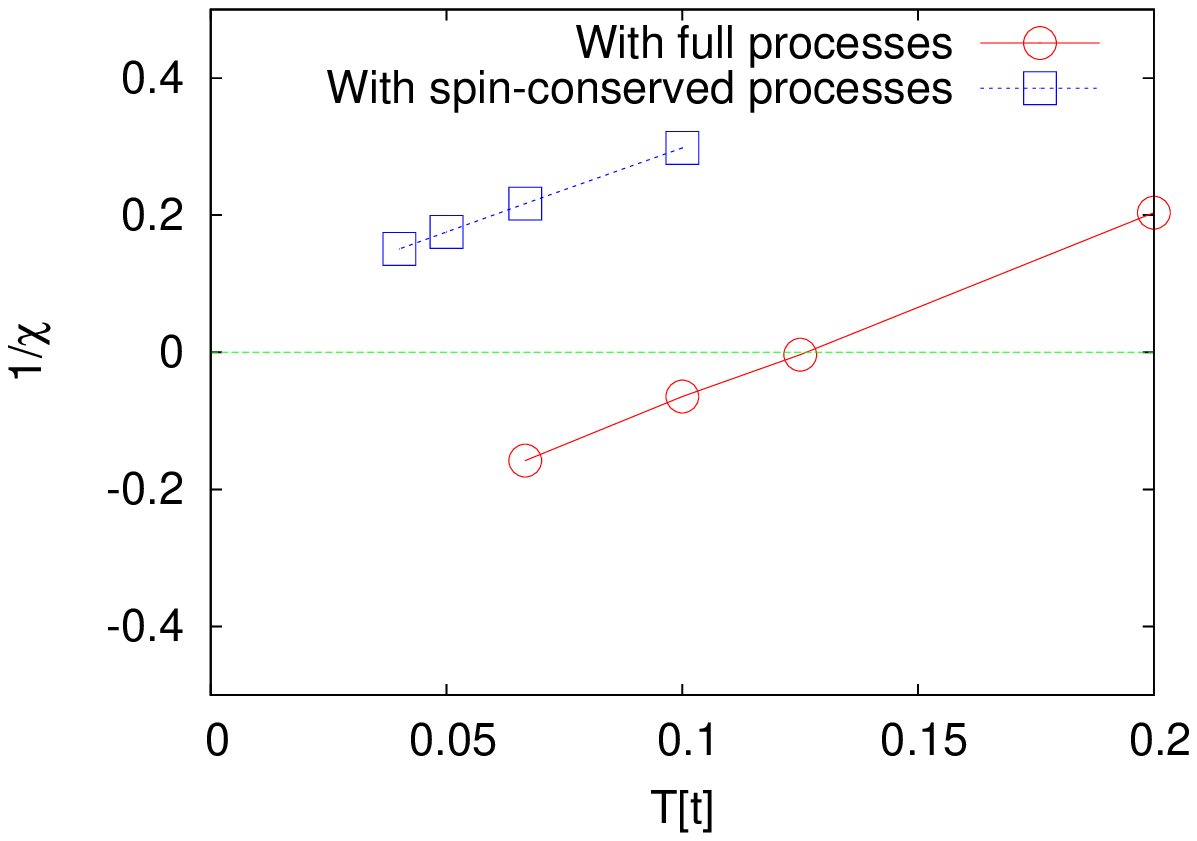}} &
      (d)\resizebox{0.4 \columnwidth}{!}{\includegraphics{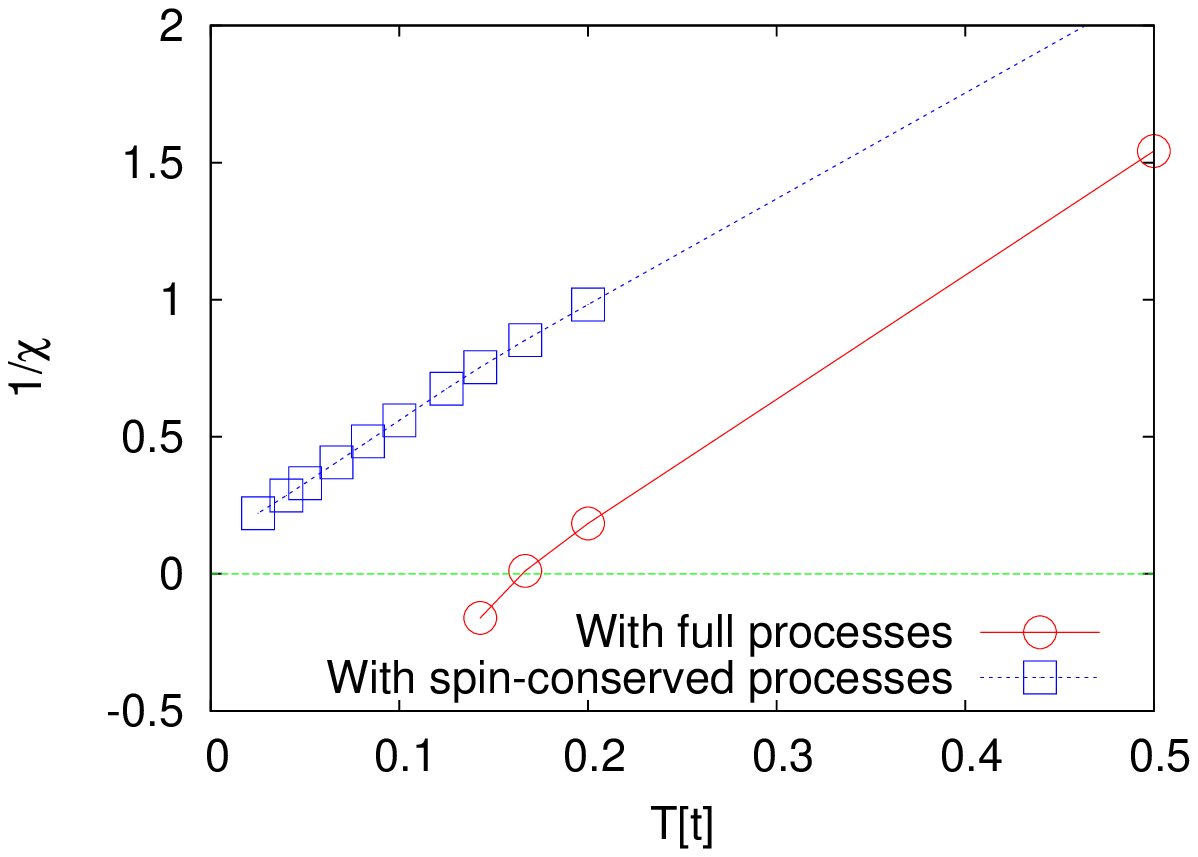}} 
    \end{tabular}
\end{center}
\caption{\label{fig:fr}(Color online) 
Sign changes of $1/\chi$ with (a) $h=0$ and $\alpha = 1$, (b) $h = 1$ and $\alpha = 0$, (d) $h = 1$ and $\alpha = 1$, and (d) $h = 1$ and $\alpha = 2$. The red circles denote the result by solving Eq.~(\ref{eq:latb}) and the blue squares denote the result by solving Eq.~(\ref{eq:latbc}). 
The filling $\nu = 1/8$ and the attractive on-site coupling $U = -3t$. 
}
\end{figure}

We show the divergences of $\chi$ in $1/\chi$ in Fig.~\ref{fig:fr}. 
We consider both cases with full and spin-conserved processes. 
Figure \ref{fig:fr}(a) shows that the critical temperature $T_{\rm c}$ with full processes is larger than 
that with spin-conserved processes in the system with $h = 0$ and $\alpha = 1$.  
Figure \ref{fig:fr}(b) shows that the results by solving both equations are same in the system without the spin orbit coupling. 
There is no critical temperature, since the Pauli depairing effect is so strong that the Cooper pairs are destroyed. 
Figures \ref{fig:fr}(c) and (d) show that the spin-flip processes are important to induce the superconducting phase. 
The spin-conserved processes can not overcome the Pauli depairing effect so that the critical temperature is zero. 


\begin{thebibliography}{99}
\bibitem{Zelse}
I. \ifmmode \check{Z}\else \v{Z}\fi{}uti\ifmmode \acute{c}\else \'{c}\fi{}, J. Fabian, and S. Das Sarma, Spintronics: Fundamentals and applications, Rev. Mod. Phys. {\bf 76}, 323 (2004). 
\bibitem{Hasan}
M. Z. Hasan and C. L. Kane, Colloquium: Topological insulators,
Rev. Mod. Phys. {\bf 82}, 3045 (2010).
\bibitem{Fulde}
P. Fulde and R. A. Ferrell, 
Superconductivity in a Strong Spin-Exchange Field, 
Phys. Rev. {\bf 135}, A550 (1964).
%
\bibitem{Larkin}
A. I. Larkin and Y. N. Ovchinnikov, 
Nonuniform state of superconductors, 
Zh. Eksp. Teor. Fiz. {\bf 47}, 1136 (1964) [Sov. Phys. JETP {\bf 20} 762 (1965)].
\bibitem{YanasePRB}
Y. Yanase, 
Angular Fulde-Ferrell-Larkin-Ovchinnikov state in cold fermion gases in a toroidal trap, 
Phys. Rev. B {\bf 80}, 220510 (2009).
%
\bibitem{Yoshida}
T. Yoshida, M. Sigrist, and Y. Yanase, 
Pair-density wave states through spin-orbit coupling in multilayer superconductors, 
Phys. Rev. B {\bf 86}, 134514 (2012).
\bibitem{Kitaev}
A. Y. Kitaev, 
Unpaired Majorana fermions in quantum wires, 
Phys. Usp. {\bf 44}, 131 (2001).
\bibitem{Nayak}
C. Nayak, S.H. Simon, A. Stern, M. Freedman, and S. D. Sarma, 
Non-Abelian anyons and topological quantum computation, 
Rev. Mod. Phys. {\bf 80}, 1083 (2008).
\bibitem{Alicea}
J. Alicea, 
New directions in the pursuit of Majorana fermions in solid state systems, 
Rep. Prog. Phys. {\bf 75}, 076501 (2012).
\bibitem{Sasaki}
S. Sasaki, M. Kriener, K. Sagawa, K. Yada, Y. Tanaka, M. Sato and Y. Ando, 
Topological Superconductivity in Cu$_{x}$Bi$_{2}$Se$_{3}$, 
Phys. Rev. Lett. {\bf 107}, 217001 (2011).
\bibitem{Fu}
L. Fu and E. Berg, 
Odd-Parity Topological Superconductors: Theory and Application to Cu$_{x}$Bi$_{2}$Se$_{3}$, 
Phys. Rev. Lett. {\bf 105}, 097001 (2010).
\bibitem{ZhangSci}
X.-L. Zhang and W.-M. Liu, Electron-Phonon Coupling and its implication for the superconducting topological insulators, 
Sci. Rep.  {\bf 5}, 8964 (2015).
\bibitem{Sau}
J. D. Sau, R. M. Lutchyn, S. Tewari, and S. D. Sarma, 
Generic New Platform for Topological Quantum Computation Using Semiconductor Heterostructures, 
Phys. Rev. Lett. {\bf 104}, 040502 (2010)
\bibitem{Sato}
M. Sato, Y. Takahashi, and S. Fujimoto, 
Non-Abelian topological orders and Majorana fermions in spin-singlet superconductors, 
Phys. Rev. B {\bf 82}, 134521 (2010).
\bibitem{Shitade}
A. Shitade and Y. Nagai, 
Orbital angular momentum in a nonchiral topological superconductor, 
Phys. Rev. B {\bf 92}, 024502 (2015).
\bibitem{NagaiJPSJ}
Y. Nagai, Y. Ota, and M. Machida,
Impurity effects in a two-dimensional topological superconductor: A link of Tc-robustness with a topological number, 
J. Phys. Soc. Jpn. {\bf 83}, 094722 (2014).
\bibitem{Georges:1996hv}
A. Georges, G. Kotliar, W. Krauth, and M.J. Rozenberg, 
Dynamical mean-field theory of strongly correlated fermion systems and the limit of infinite dimensions, 
Rev. Mod. Phys. {\bf 68}, 13 (1996).
\bibitem{Werner}
P. Werner and A. J. Mills, 
Hybridization expansion impurity solver: General formulation and application to Kondo lattice and two-orbital models, 
Phys. Rev. B {\bf 74}, 155107 (2006).
\bibitem{Gull}
E. Gull, A. J. Millis, A. I. Lichtenstein, A. N. Rubtsov, M. Troyer, and P. Werner, 
Continuous-time Monte Carlo methods for quantum impurity models, 
Rev. Mod. Phys, {\bf 83}, 349 (2011).
\bibitem{Micnas:1990ee} 
R. Micnas, J. Ranninger, and S. Robaszkiewicz, 
Superconductivity in narrow-band systems with local nonretarded attractive interactions, 
Rev. Mod. Phys. {\bf 62}, 113 (1990).
\bibitem{Matetskiy}
A.V. Matetskiy, S. Ichinokura, L.V. Bondarenko, A.Y. Tupchaya, D.V. Gruznev, A.V. Zotov, A.A. Saranin, R. Hobara, A. Takayama, and S. Hasegawa, 
Two-Dimensional Superconductor with a Giant Rashba Effect: One-Atom-Layer Tl-Pb Compound on Si(111), 
Phys. Rev. Lett. {\bf 115}, 147003 (2015).
\bibitem{Li}
L. J. Li, E. C.T. O'Farrell, K. P. Loh, G. Eda, B. \"{O}zyilmaz, and A. H. Castro Neto, 
Controlling many-body states by the electric-field effect in a two-dimensional material, 
Nature (2015) doi:10.1038/nature16175.
\bibitem{Thouless}
D. J. Thouless, M. Kohmoto, M. P. Nightingale, and M. den Nijs, 
Quantized Hall Conductance in a Two-Dimensional Periodic Potential, 
Phys. Rev. Lett. {\bf 49}, 405 (1982).
\bibitem{Kohmoto}
M. Kohmoto, 
Topological invariant and the quantization of the Hall conductance, 
Ann. Phys. {\bf 160}, 343 (1985).
\bibitem{Haule}
K. Haule, 
Quantum Monte Carlo impurity solver for cluster dynamical mean-field theory and electronic structure calculations with adjustable cluster base, 
Phys. Rev. B {\bf 75}, 155113 (2007).
\bibitem{WernerPRL}
P. Werner, A. Comanac, L.Medici, M. Troyer, and A. J. Millis, 
Continuous-Time Solver for Quantum Impurity Models, 
Phys. Rev. Lett. {\bf 97}, 076405 (2006). 
\bibitem{note}
As $|\omega_{n}| \rightarrow \infty$, local Green's functions in the original and effective impurity models are $\hat{G}^{\rm loc}(i \omega_{n}) \sim 1/(i \omega_{n}) + (\sum_{\Vec{k}} \hat{h}_{0}(\Vec{k})+ \hat{\Sigma}_{0})/(i \omega_{n})^{2}$ 
and $\hat{G}_{\rm f}(i \omega_{n}) \sim 1/(i \omega_{n}) + (\hat{H}_{\rm f}+ \hat{\Sigma}_{0})/(i \omega_{n})^{2}$, respectively, with the zero-th order self-energy $\hat{\Sigma}_{0}$.
Thus, the self-consistent condition in the DMFT, $\hat{G}^{\rm loc}(i \omega_{n}) = \hat{G}_{\rm f}(i \omega_{n})$, leads to Eq.~(\ref{eq:hf}).
\bibitem{note2}
Local Green's function is $\hat{G}^{\rm loc}(i \omega_{n}) = a \hat{1} + b \hat{\sigma}_{3}$ when the self-energy is diagonal in spin space. 
The hybridization function in the ct-HYB becomes diagonal in the this case. 
\bibitem{iQist}
Li Huang, Yilin Wang, Zi Yang Meng, Liang Du, Philipp Werner and Xi Dai, 
iQIST: An open source continuous-time quantum Monte Carlo impurity solver toolkit, 
Comp. Phys. Comm. {\bf 195}, 140 (2015).
\bibitem{Hoshino}
S. Hoshino and P. Werner, P
Superconductivity from emerging magnetic moments, 
Phys. Rev. Lett. {\bf 115}, 247001 (2015).
\bibitem{Freericks}
J.K. Freericks, M. Jarrell and D.J. Scalapino, 
Holstein model in infinite dimensions, 
Phys. Rev. B {\bf 48}, 6302 (1993).






\bibitem{suppl}
See Supplemental materials for the detail arguments on our DMFT calculations.

\bibitem{Xu}
Y. Xu and C. Zhang, 
Berezinskii-Kosterlitz-Thouless Phase Transition in 2D Spin-Orbit-Coupled Fulde-Ferrell Superfluids,
Phys. Rev. Lett. {\bf 114}, 110401 (2015).
\bibitem{note3}
Solving the mean-field BCS linearized gap equations, we have $T_{\rm c} \sim T_{\rm c}^{0} \exp \left[\frac{1}{|U|} \left[ \frac{1}{N^{0}} - \frac{1}{N(\alpha,h)} 
\right] \right]$ if $\alpha \gg h$. 
Here, $T_{\rm c}^{0}$ and $N^{0}$ are a critical temperature and a density of states at the Fermi energy for $h = 0$ and $\alpha = 0$, respectively. 
The weak-coupling formula indicates that $T_{\rm c}$ is subjected to the density of states on the Fermi surfaces $N(\alpha,h)$, which monotonically increases as $\alpha$ when $\nu$ is fixed. 
%
\bibitem{Devreese}
J. P. A. Devreese, J. Tempere, and C. A. R. Melo, 
Effects of Spin-Orbit Coupling on the Berezinskii-Kosterlitz-Thouless Transition and the Vortex-Antivortex Structure in Two-Dimensional Fermi Gases, 
Phys. Rev. Lett. {\bf 113}, 165304 (2014).

\bibitem{Wang:2012fjc}
Z. Wang and S.-C. Zhang, 
Simplified Topological Invariants for Interacting Insulators, 
Phys. Rev. X {\bf 2}, 031008 (2012).



\bibitem{NagaiRapid}
Y. Nagai, 
Robust superconductivity with nodes in the superconducting topological insulator Cu$_{x}$Bi$_{2}$Se$_{3}$: Zeeman orbital field and nonmagnetic impurities, 
Phys. Rev. B {\bf 91}, 060502(R) 2015. 


\end{thebibliography}
\end{document}